# Helium load on W-O coatings grown by pulsed laser deposition

R. Mateus[1,*], D. Dellasega[2,3], M. Passoni[2,3], Z. Siketić[4],
I. Bogdanović Radović[4], A. Hakola[5], E. Alves[1]

[1]*Instituto de Plasmas e Fusão Nuclear, Instituto Superior Técnico, Universidade de Lisboa, Av. Rovisco Pais, 1049-001 Lisboa, Portugal*

[2]*Dipartimento di Energia, Politecnico di Milano, via Ponzio 34/3, 20133 Milano, Italy*

[3]*IFP, CNR, via R. Cozzi 53, Milano 20125, Italy*

[4]*Ruđer Bošković Institute, P.O. Box 180, 10002 Zagreb, Croatia*

[5]*VTT Technical Research Centre of Finland Ltd., Finland*

*corresponding author: rmateus@ipfn.ist.utl.pt

**Abstract**

W-O deposits with complex morphologies and significant He contents will growth on the surface of plasma facing components exposed to He discharges. To mimic the re/co-deposition process, W-O coatings were loaded with He by implanting $He^+$ ions on W films grown by pulsed laser deposition (PLD). The use of appropriate PLD experimental parameters such as pressurised Ar or He background atmospheres induces the deposition of porous or compact W structures enhancing afterwards the gathering of different amounts of O under exposure to atmospheric air. After multiple ion implantation stages using 150 keV, 100 keV and 50 keV incident $He^+$ ion beams with a total fluence of $5 \times 10^{17}$ ion/cm$^2$, significant amounts of He were identified in porous coatings by Rutherford backscattering (RBS). Time-of-flight elastic recoil detection (ToF-ERDA) measurements showed that most of the implanted He was already released from the porous coatings five month after implantation, while for the case of compact layers the He content remains significant at deeper layers and smoothly decrease towards the surface, as result of a different morphology and nanostructure. The proposed method involving PLD and ion implantation seems adequate to produce W-O films load by He that may be used as reference samples for fusion investigations.

## 1. Introduction

During the operation of fusion devices the plasma will strongly interact with plasma facing components (PFC) leading to the erosion of the exposed surfaces, co-deposition of eroded material or environment impurities as oxygen (O), and trapping of fuel as well as isotopic products (He-4) of the D-T reaction [1]. Tungsten (W) is the chosen material for the divertor of ITER [1] and the use of full W PFC walls and divertors is under investigation in different plasma devices [2,3]. Experimental results revealed that the grown deposits may present distinct microstructures and compositions with various impurities. This is the case of compact W-O layers covering optical mirrors located at the first wall and divertor of the ASDEX Upgrade tokamak [2], and of porous W-O deposits found on the outer divertor strike point tiles of the same device [3]. The effects caused by deuterium and helium plasma exposure are related to the different W-O structures. There is therefore need to determine the outcomes arising from the modifications in dedicated experiments to predict the properties and lifetimes of the irradiated PFCs. To this purpose, W-O coatings loaded with deuterium or helium (He) and resembling re-deposited layers on PFCs in W-based fusion devices are being produced under the EUROfusion consortium [4-6]. The present work reports part of the activities aiming the production of W-O coatings loaded by He in order to develop W(He)-based materials by ion implantation similar to deposits observed in tokamaks. Two different as-deposited coatings were used in the experiment: porous-W with an open nanocrystalline structure and also compact-W presenting poor crystallinity. In order to prevent the occurrence of significant morphological changes enhancing gas release, the use of a moderate irradiation dose for He ions was established. The investigation of He loading and release dynamics in the two microstructures, also after long time storage, was carried out by ion beam analysis.

## 2. Experimental procedure

Pulsed Laser Deposition (PLD) technique has been exploited to produce appropriate coatings for the research using a pure metallic W target as W source and a proper background pressure of argon (Ar) or He. The role of different PLD parameters in the deposition procedure, such as the wavelength of the laser beam used to ablate the W source (532 nm in this case), the laser fluence per pulse (close to 15 J/cm$^2$), the target-to-substrate distance (7 cm) or the pressure and composition of the used background gas are detailed elsewhere [4,5]. PLD parameters are properly tailored to produce porous-W and compact-W coating that mimic W-O re-deposited microstructures observed in tokamaks [2,3] after O absorption [5]. In the present experiment, porous-W coatings were prepared by using a 50 Pa Ar atmosphere, promoting the porous morphology characterized by the presence of nano-grains disposed along the columnar growth of the films. For the compact-W deposition, a 70 Pa atmosphere of He, the same PLD geometry and similar laser pulses to ablate the W target were used. The compact morphology is only characterized by the growth of nano-grains spread out along the entire films structure. The gathering of O through the W lattice occurs afterwards by exposing the as-deposited samples to atmospheric air [4,5]. The coatings were deposited on stainless steel (SS) 316L substrates. In order to reinforce the adhesion of the coatings to the substrates a pure and thin 150 nm crystalline W interface was deposited initially in vacuum ($10^{-3}$ Pa). Table 1 resumes some of the relevant parameters used to growth both porous and compact W coatings and crystalline W interfaces. Recent experiments revealed the absence of Ar or He on the coatings although the PLD depositions are carried out under Ar or He background atmospheres [4,5] (section 3.3).

A Zeiss Supra 40 field emission scanning electron microscope assisted with an X-ray detector was used to characterize surface morphologies by scanning electron microscopy (SEM) and to quantify the elemental W and O contents by energy dispersion X-ray spectroscopy (EDX). X-ray diffraction (XRD) measurements were performed with a Panalytical X'Pert PRO X-ray diffractometer in the θ-2θ configuration using a copper (Cu) X-ray source for phase identification.

A 210 kV high flux ion implanter was used to irradiate simultaneously all the porous and compact coatings with normal incidence of $^4He^+$ ion beams. Aiming to prevent a possible saturation of He on the W lattices, the final depth ranges were spread out and diluted through the implantation zone by using three different implantation stages with a sequential decrease in the energy of incident ions. The corresponding ion beam energies and fluences of the first, second and third stages were, respectively, 150 keV and $2 \times 10^{17}$ ion/cm$^2$, 100 keV and $2 \times 10^{17}$ ion/cm$^2$, and 50 keV and $1 \times 10^{17}$ ion/cm$^2$, being the total dose equal to $5 \times 10^{17}$ ion/cm$^2$. Current densities of the individual beams of 4.3 µm/cm$^2$ assure that the irradiations of the metallic samples were performed at room temperature (RT). The depth ranges of ions within the materials were evaluated by using the SRIM code [7].

The thickness and the W and O elemental depth profiles of the as-deposited and as-implanted samples were analysed by Rutherford backscattering (RBS, with 2 MeV He$^+$) and by elastic backscattering spectrometries (EBS, with 1.5 MeV H$^+$, where Gurbich's evaluated non-Rutherford cross-sections for O are about 2.5 times higher than the Rutherford cross-section at the same energy [8]) making use of the NDF code [9] enabling simultaneous analysis involving different ion beam techniques. The analysis took into consideration the main composition of the SS substrates. He retained amounts could be evaluated indirectly by RBS from the decrease of the backscattering yields of heavy W in the implanted samples. Nevertheless and particularly for a direct quantification of He, time-of-flight elastic recoil detection (ToF-ERDA) experiments using 23 MeV $^{127}I^{6+}$ ion beams at a glancing angle of 20° were also performed to evaluate the release of He and to quantify the depth profiles of a large number of elements along the superficial layers. At the end, O amounts evaluated from SEM-EDX, EBS and ToF-ERDA could be compared.

A first elemental depth profiling quantification was performed from the time-of-flight data with the POTKU software [10], which allows checking for selective composition changes during spectra collection. This is particularly useful to evaluate possible losses of (weakly bounded) helium during ion bombardment. As a second stage of the analysis, the compositions achieved from the POTKU code were used as input files of the Monte Carlo CORTEO software [11], which takes into account all the energy spread contributions that may affect the final elemental depth quantifications as multiple scattering events and energy straggling, and a fine tuning of the elemental POTKU results was obtained. The procedure is particularly useful to check the depth profiling of heavy elements as W.

## 3. Results and discussion

### 3.1 Characterization of porous and compact morphologies

As evidenced from the secondary electron SEM images of Fig. 1, two distinct W microstructures were formed by tuning the PLD deposition parameters. From the top view of a porous-W film (a) we observe the cauliflower morphology, which commonly is associated to a porous and columnar growth involving the individual crystallites. In opposition, any contrast is observed in the top view of a compact-W film (b), which is associated to a smooth, compact and featureless growth. The corresponding cross-section images confirm the presence of the two distinct morphologies: the superficial cauliflower morphology of compact-W corresponds to the growth of nanostructured columns, 100 nm wide, along the normal direction to the substrate (c); the cross-section view of compact-W reveals the same featureless growth already observed in the top view image. Finally, and for the present depositions, a crystalline W adhesion layer on the interface between an SS substrate and a porous coating is detailed by a magnified cross-section view image (e).

It is possible to characterize the crystallinity of the deposited films from the XRD diffractograms in Fig. 2. The porous film exhibits a quite broad reflection centered at 39.8 degrees, evidencing the growth of a bcc W-based microstructure. Nevertheless, the reported value for the (110) diffraction peak of pure bcc W is about 40.3° [12]. From the Bragg's law, and considering the wavelength of 0.154 nm for the Cu-K$\alpha_1$ line of the X-ray source, a decrease in the (110) peak close to 0.50 degrees corresponds to an increment of the lattice parameter $a$ of the bcc W phase from the reference value of 0.317 nm to about 0.320 nm, and to a lattice expansion $\Delta a/a$ of about 1.2% relative to pure bcc W. The structural swelling is easily justified by addition of O and other impurities in the W unit cell [13,14]. Considering the broadening of the (110) peak, the mean crystallite domain size calculated by using the Scherrer equation is about 7 nm [5,13]. The bcc W structure in the porous-W coatings is also confirmed by the corresponding diffraction peaks detected with lower intensities for planes (200), (112) and (220). On the other side, the diffractogram of compact-W shows only a wide and broad band typically centered, once again, at the diffraction peak for plane (110) of bcc W. From the Scherrer equation the mean crystalline size of the grains is lower than 2 nm [5,13]. These data show that the porous samples exhibit higher crystallinity in respect to the compact ones. This may be related to the formation of crystalline nanoparticles during the ablation process due to the interaction with the Ar atmosphere.

The use of an Ar background atmosphere results in a significant decrease of the kinetic energy of the ablated W atoms impinging on the substrate due to the scattering imposed by the Ar atoms, thus leading to the growth of non-compact but porous structures, as observed in Fig. 1(a) and (c). In opposition to Ar atmospheres, He has been chosen as the background gas for the formation of compact-W coatings because it is the noble gas with the smallest atomic mass and consequently with the lowest thermalizing effect on the ablated atoms. An earlier investigation has shown a structural transition from a crystalline to a highly compact morphology growth in the deposits at about 30 Pa by varying the He background pressure in the range from $10^{-3}$ Pa to 200 Pa [5], and by raising the He pressure the deposits appear to be formed by even smaller grains, with a corresponding decrease in films density and rise of bcc W lattice parameters. Also the O content may rise quickly from 5% at. to 20 at.%. within a slight pressure change from 40 Pa to 50 Pa. A larger retention of O may be related to the formation of new kinds of defects and energetic traps in the films. Nevertheless, the final O content (15%-20%) is always lower compared to the case of porous deposits (~50%) [5]. High surface-to-volume ratio of these microstructures, enhanced by a columnar growth, induces the spontaneous gathering of O after deposition in the coatings during exposition to ambient air [4]. The O content, as quantified by EDS in the surface of the porous-W films using in the present experiment was about 55 at.% [4]. Instead, thanks to its compact morphology, compact-W films showed O contents close to 16 at.% [5]. Different experiments demonstrated that O may be absorbed after W deposition without reacting chemically [15] and this is the case of the present samples [16].

As a consequence of the PLD deposition procedure, and for a better characterization of the samples before implantation, also the phase of the crystallites identified by XRD, and the diluted/trapped O contents present in the surface of the samples and evaluated by EDX, were included in Table 1.

**3.2 Multiple implantation stages of helium**

Earlier implantation experiments with energetic He ions evidenced drastic changes to the surface morphology of polycrystalline W maintained in the 500-900 ºC temperature range when the He dose exceeded 1 x $10^{18}$ ion/cm$^2$, promoting the release of He through nanoscale channels by the coalescence and movement of He bubbles through the implantation zone [17]. As referred in section 2, the present irradiation campaign was performed with a total He$^+$ dose of 5 x $10^{17}$ ion/cm$^2$ and keeping the samples at RT. The main idea of the procedure was to avoid major morphological changes that could enhance the degassing of He [17]. Depth ranges and path length distribution of implanted He were calculated with the SRIM code considering materials with very distinct O contents [7]. The stopping power imposed by W-O matrixes to the incident ions and quantified with a linear combination of the elemental stopping powers by the Bragg's rule, giving rise to very accurate results for the case of heavy elements as O and W [7]. Once the irradiations of the coatings were performed simultaneously, lower energy losses imposed to the incident ions and higher ion ranges are predicted within the porous structures, while they present matrixes with lower average Z-values [7,18]. Nevertheless, different W:O ratios could be present along the depth range of the impinging ions. Also the calculations do not take into consideration the crystalline structure of the target or the role of columnar or grain boundaries and lattice defects [7,18]. Not all the ions implanted with an unique energy take the same depth due to the energy spread of ions in matter, and by calculating the average projected range, Rp, the corresponding standard deviation, ΔRp, and by knowing the implanted ion dose, Gaussian-like depth profile ranges and maximum concentrations of the implanted species are predicted [7,18]. Table 2 presents Rp and ΔRp depth ranges calculated for two distinct W:O compositions: pure W and the W:O (37:62) stoichiometry, the composition quantified by EBS/RBS for the as-deposited porous films (section 3.3). Considering the corresponding depth ranges and ion fluences for the three individual irradiation stages (performed with 150 keV He$^+$, 100 keV He$^+$ and 50 keV He$^+$), total depth profiles for implanted ions in pure W and in the present porous-W are obtained (see Fig. 3). For the porous films, maximum He contents of about 15 at.% were predicted (Fig. 3). Outcomes arising from lattice defects will lead to higher retentions of He close to the surface and to lower He depth ranges. Since the O amount in compact layers is lower than in porous ones, it is predicted before the irradiation campaign that the impinging ions will not reach the SS substrate or the W interface whatever the coating structure, porous or compact.

### 3.3 Helium load and release dynamics on porous and compact coatings

Composition of porous and compact W films has been checked after the deposition procedure (as-deposited), right after He implantation (as-implanted) and five months after implantation (aged). As-deposited films have been characterized by SEM, EDS, EBS, RBS and ToF-ERDA, as-implanted films have been analysed by EBS and RBS and aged films have been characterized by EBS, RBS and ToF-ERDA. Additionally, and with the purpose to confirm the origin of retained He, additional ToF-ERDA measurements had been carried out to new sets of porous and compact films deposited by PLD at the same experimental parameters.

Fig. 4.a shows the EBS spectra collected from the porous coatings before and after implantation, where the vertical arrows indicate the energies relative to the presence of W and O at the surface layer. From the elastic scattering of incident protons by $^{16}$O, $^{16}$O(p,p)$^{16}$O, significant amounts of O on porous-W are quantified. The same behaviour is also visible from the decrease of the backscattering yield of heavier W along the coating's depth. An interface region with pure W (identified as the crystalline W adhesion layer in Fig. 1) is located between the film and the SS substrate. In comparison to the incidence of protons, the energy loss of He ions in matter is higher, since it depends on the square of the incident ion charge [7,18]. Therefore we are highly sensitive to changes in the superficial depth profile of heavy W by RBS (Fig. 4.b), where an expected decrease in the backscattering W yield after the implantation procedure is observed, being the behaviour explained by the addition of the He impurity [19,20]. Therefore and despite the absence

of a He yield, a retained depth profile for He may be quantified by RBS [19,20]. The complementary EBS/RBS analyses in the as-deposited porous films point to an O content close to 62.5 at.% and to coating's thickness of about 112 x $10^{17}$at/cm$^2$, followed by a pure W interface close to 9.8 x $10^{17}$at/cm$^2$ (~155 nm considering bulk W density). After implantation the coating's thickness decreases to 107 x $10^{17}$at/cm$^2$ due to physical sputtering and a maximum He content of 23 at.% is evaluated. Taking into consideration the total fluence of He implantation, 5 x $10^{17}$ at/cm$^2$, it is predicted large sputtering yield of 0.94 caused by impinging energetic He$^+$ on porous-W. Exceptionally large sputtering yields caused by the incidence of heavy energetic ions on tungsten oxide were observed before [21]. The behaviour need to be studied with further experiments.

In opposition to the case of porous layers, and due to the compact nature of the coatings and possibly, due to a lower O content, the erosion by physical sputtering did not occur significantly by irradiating compact-W. Moreover, and if a low O content exist, the analysis to EBS (Fig. 4(c) ) and RBS spectra (not presented in Fig. 4) are not sensitive to quantify the existing amount. Additional differences in the elemental profiles before and after implantation are explained by changes in the energy resolution of the detector.

Fig. 5 presents the depth profiles for He, O and W quantified from the EBS and RBS spectra of the porous coatings collected before (a) and after helium implantation (b). As predicted by SRIM, the maximum retained He amount did not occur at the surface, being the implanted depth zone in the range from 0.5 to 2 x $10^{18}$ at/cm$^2$ (see Fig. 5(b)). SRIM simulations pointed to an implanted depth zone in the depth range from 1.0 to 4.5 x $10^{18}$ at/cm$^2$ with a lower maximum retained He amount (15 at.% vs. 23 at.%) (see Fig. 3)). The results seems to agree with the occurrence of a huge physical sputtering [21] in the porous surfaces during irradiation and with a fast retention of He atoms along lattice defects, preferentially, in columnar and grain boundaries of the porous bcc structure [17,22]. In face of the present results, the formation of He bubbles at the present irradiation parameters needs to be investigated.

Ar and He are noble gases, thus insoluble in metals and migrate rapidly through metallic lattices. The trapping in lattice defects or a degassing at the surface should be quite fast [17,22] even at room temperature. Due to the grazing geometry of the analysis, ToF-ERDA is really sensitive to the presence of the individual elements nearby the superficial layers. Very recent ToF-ERDA measurements carried out in new and as-deposited W-O films deposited at PLD parameters similar to those ones of Table 1 revealed the superficial elemental contents of both porous-W (15 at.% H, 1.9 at.% C, 2.5 at.% N, 60 at.% O, 20.6 at.% W) and compact-W coatings (2.1 at.% H, 0.7 at.% C, 2.0 at.% N, 25 at.% O, 70 at.% W). Higher contents for H, C, N and O in porous surfaces agree with an easier trapping behaviour of environmental contaminants in a porous structure [22]. As example of the analysis, Fig. 6(a) presents the coincident map collected from an as-deposited compact-W sample obtained with the particle (ERDA) detector and the two time-of-flight (ToF) detectors, where all the elemental yields are well separated (the yield for I relates with the primary beam). As it was referred before, and despite the use of a pressurize He atmosphere during deposition, He is not present in the as-deposited sample and it it only incorporated under ion implantation.

Although a huge He retention in as-implanted porous-W coatings has been observed by RBS (maximum of 23 at.%; Fig. 4.b), ToF-ERDA maps confirmed that most of the retained He has been released already from aged porous-W five months after implantation (data not shown), while a low 3.1 at.%. average He concentration was quantified along a superficial depth close to $17.0 \times 10^{17}$ at/cm$^2$. The fast degassing is justified in face of a porous structure acting as a path way through the coatings depth [17,22]. The remaining detected elements in the same aged coating were H (7.4 at.%), C (4.2 at.%), N (3.0 at.), O (56 at.%) and W (26 at.%). Nevertheless, the equivalent ToF-ERDA map of Fig. 6(b) collected from an aged compact-W coating confirmed that a significant amount of He (13.8 at.%) remains within a depth layer of $12.0 \times 10^{17}$ at/cm$^2$ five month after irradiation, being the remaining elemental composition H (2.3 at.%), C (2.0 at.%), N (1.3 at.%), O (15 at.%) and W (65 at.%). Fig. 7 shows the quantified elemental depth profiles for W, O and He in the aged compact-W achieved from the POTKU [10] and CORDEO [11] codes. In particular, it is observed a smooth increase in the retained He amounts at deeper depths, being the result compatible with the release of He at the surface, while significant He amounts remain present at deeper depths. Confirming the result, we may affirm that there was no loss of light elements as helium caused by ion beam damage during the ToF-ERDA measurement [10], while only a statistical fluctuation of the light elements concentration was observed over time. Also the contributions arising from energy spread effects as multiple scattering [11] are not sufficient to justify the decrease of W yields at deeper depths.

X-ray induced emission is only sensitive for the analysis of superficial O. EBS/RBS is truly accurate to quantify significant amounts of O along the entire depth of the film and ToF-ERDA is extremely sensitive for the analysis of the superficial layers. The O contents measured at three different research laboratories by EDX, EBS/RBS and ToF-ERDA for the as-deposited porous (54 at.%, 62 at.% and 60 at.%, respectively) and for the compact coatings (16 at.%, not sensitive by RBS, and 25 at.%, respectively) are in impressive agreement, signalising the accuracy of the global analyses. All the campaig points ion implantation as a promising technique to load He in W films produced by PLD. Nevertheless, the lifetime of enriched He films will extremely depend of the internal microstructures and ion beam energies, implantation dose and current densities implemented during irradiation.

4. Summary

W-O coatings were prepared by PLD to mimic the load of He on W-O deposits in tokamaks. The use of Ar and He atmosphere as background gas and tuned PLD deposition parameters induced the deposition of porous and compact coatings avoiding the retention of the background gas elements. He load was only obtained after ion implantation carried out with sequential implantation stages involving 150 keV, 100 keV and 50 keV He$^+$ ion beams, showing that the procedure is suitable to produce W-O coatings containing He that may be used standards for fusion investigations. Film composition has been assessed by EDX, EBS/RBS and ToF-ERDA showing a substantial agreement between the different techniques. EBS and RBS are not sensitive to identify smooth changes in the depth profiles of retained He as observed from structures. Conversely ToF-EDA has been demonstrated to be a reliable technique for depth ranges of the order of 1500 x $10^{17}$ at/cm$^2$.

He retention dynamics is deeply affected by film morphology, nanostructure and composition. The results evidence that a columnar morphology allow retaining significant He amounts. Nevertheless, also a fast release takes place, and five months after implantation, most of the He amount has been released from the surfaces. Significant erosion of the porous surface is predicted to occur under irradiation with energetic ions. The compact morphology is able to retain higher amounts of He five months after exposure, being the result in agreement with a smooth degassing of He and with the retention of significant He amounts at deeper depths. Additional experiments should be carried out to characterize the He load and optimize the PLD and ion implantation procedures to produce standards of W-O loaded by He.

**Acknowledgements**

This work has been carried out within the framework of the EUROfusion Consortium and has received funding from the Euratom research and training programme 2014-2018 under grant agreement No 633053. The activity was performed in the scope of the WP PFC programme. IST also received financial support from "Fundação para a Ciência e a Tecnologia" through project UID/FIS/50010/2013. The views and opinions expressed herein do not necessarily reflect those of the European Commission.

**References**

[1] Kleyn A.W., Cardozo N.J.L., Samm U., Phys. Chem. Chem. Phys., 8, 1761-1774 (2006).

[2] Litnovsky A. et al., Nucl. Fusion, 53, 073033 (20133).

[3] Rasinski M. et al., Fusion Eng. Des., 86, 1753–1756 (2011).

[4] Maffini A., Uccello A., Dellasega D., Passoni M., Nucl. Fusion, 56, 086008 (2016).

[5] Dellasega D., Merlo G., Conti C, Bottani C.E., Passoni M., J. Appl. Phys., 112, 084328 (2012).

[6] Brezinsek S. et al., Nucl. Fusion 57, 116041 (2017).

[7] Ziegler J.F., Ziegler M.D., Biersack J.P., Nucl. Instrum. Methods Phys. Res. B, 268, 1818-1823 (2010).

[8] Gurbich A.F., Nucl. Instrum. Methods Phys. Res. B, 129, 311-316 (1997).

[9] Barradas N.P., C. Jeynes, R.P. Webb, Appl. Phys. Lett., 71, 291-293 (1997).

[10] Arstila K. et al., Nucl. Instrum. Methods Phys. Res. B, 331, 34-41 (2014).

[11] Schiettekatte F., Nucl. Instrum. Methods Phys. Res. B, 266, 1880-1885 (2008).

[12] Giorgi A.L., Physica B, 135, 420-422 (1985).

[13] Cullity B.D., Elements of X-ray Diffraction, second ed., Addison-Wesley, Reading, Massachusetts, USA, 1978.

[14] Mateus R., Sequeira M.C., Porosnicu C., Lungu C.P., Hakola A., Alves E., Nucl. Mater. Energy, 12, 462-467 (2017).

[15] Linsmeier Ch., Wanner J., Surf. Sci. 454–456, 305 (2000).

[16] Dellasega D. et al., Nanotechnology 26, 365601 (2015).

[17] Zenobia S.J., Garrison L.M., Kulcinski G.L., J. Nucl. Mater., 425, 83-92 (2012).

[18] Ryssel H. and Ruge I., Ion implantation, Wiley, Chichester, 1986.

[19] Mateus R. et al., J. Nucl. Mater., 442, S251-S255 (2013).

[20] Dias M. et al., J. Nucl. Mater., 442, 69-74 (2013).

[21] Matsunami N., Sataka M., Okayasu S., Kakiuchida H., Nucl. Instrum. Methods Phys. Res. B, 268, 3167-3170 (2010).

[22] Sharafat S. et al., J. Nucl. Mater., 347, 217-243 (2005).

**Table 1**

Relevant PLD parameters used to grow porous and compact W coatings and pure W adhesion layers, identified phases by XRD and O contents quantified by EDS in the as-deposited coatings.

| Parameter | Porous-W | compact-W | W adhesion layer |
|---|---|---|---|
| backgroud gas | Ar | He | - |
| gas pressure | 50 Pa | 70 Pa | $10^{-3}$ Pa |
| film thickness | 1 µm | 1 µm | 0.15 µm |
| phase | bcc W | bcc W | bcc W |
| O (at.%) | 54.3±5 | 16.1±2 | - |

**Table 2**

Depth ranges of incident $He^+$ ions on pure W and on W:O (37:62) simulated by SRIM.

| incident beam | pure W | | W:O (37:62) | |
|---|---|---|---|---|
| | Rp (nm) | ΔRp (nm) | Rp (nm) | ΔRp (nm) |
| 150 keV $D^+$ | 287 | 99 | 556 | 166 |
| 100 keV $D^+$ | 207 | 81 | 406 | 139 |
| 50 keV $D^+$ | 119 | 54 | 233 | 95 |

**Figure Captions**

Fig. 1. SEM images of W-O coatings: top views for porous (a) and nanostructured (b) coatings and corresponding cross-section views in (c) and (d); detail of a cross-section view of a porous coating (e).

Fig. 2. XRD diffractograms collected from both p-W and a-W coatings in the θ-2θ configuration using a Cu X-ray source.

Fig. 3. Simulated depth profiles of incident He ions on W:O (37:62) by SRIM.

Fig. 4. EBS (a) and RBS spectra (b) of p-W coatings and EBS spectra of nanostructured coatings (c) collected before and after $He^+$ implantation. The EBS and the RBS experiments were carried out with 1500 keV $H^+$ and 2000 keV $He^+$ incident ion beams, respectively, at a scattering angle of 165º.

Fig. 5. W, O and He depth profiles in as-deposited (a) and in as-implanted (b) porous-W coatings as quantified by EBS/RBS.

Fig. 6. ToF-ERDA overlapped coincidence maps of an as-deposited a-W (a) and of the implanted a-W (b) coating. The spectra were carried at a glancing angle of 20° made use of incident 23 MeV $^{127}I^{6+}$ ion beams.

Fig.7. W, O and He depth profiles in implanted a-W quantified with both POTKU and CORTEO codes.

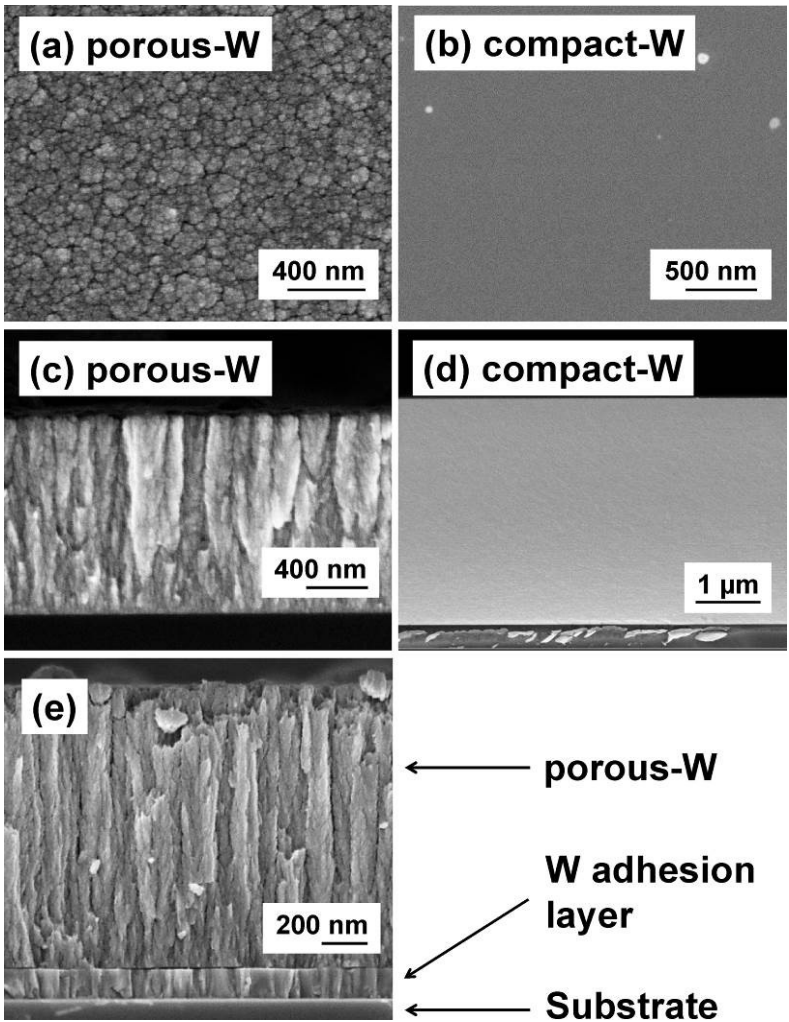

**Fig. 1**

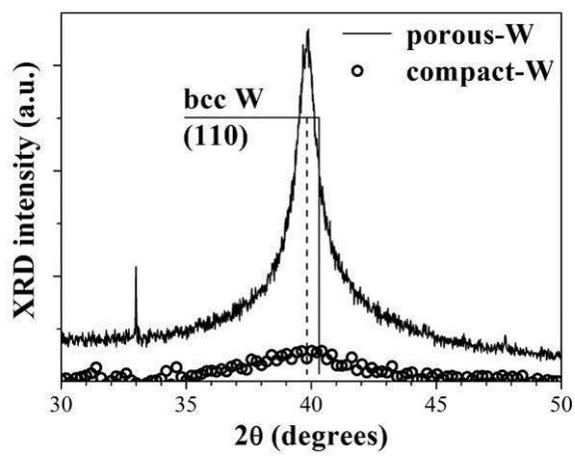

**Fig. 2**

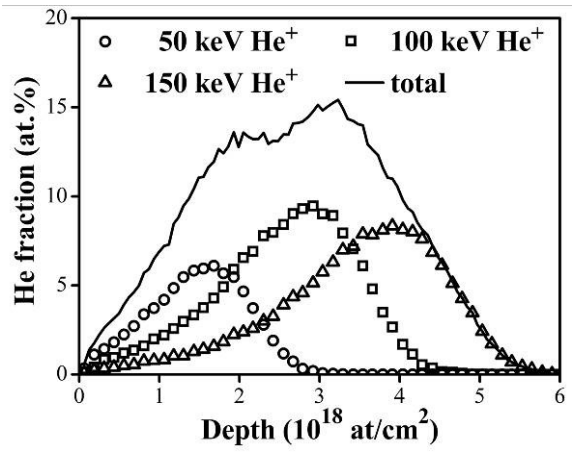

**Fig. 3**

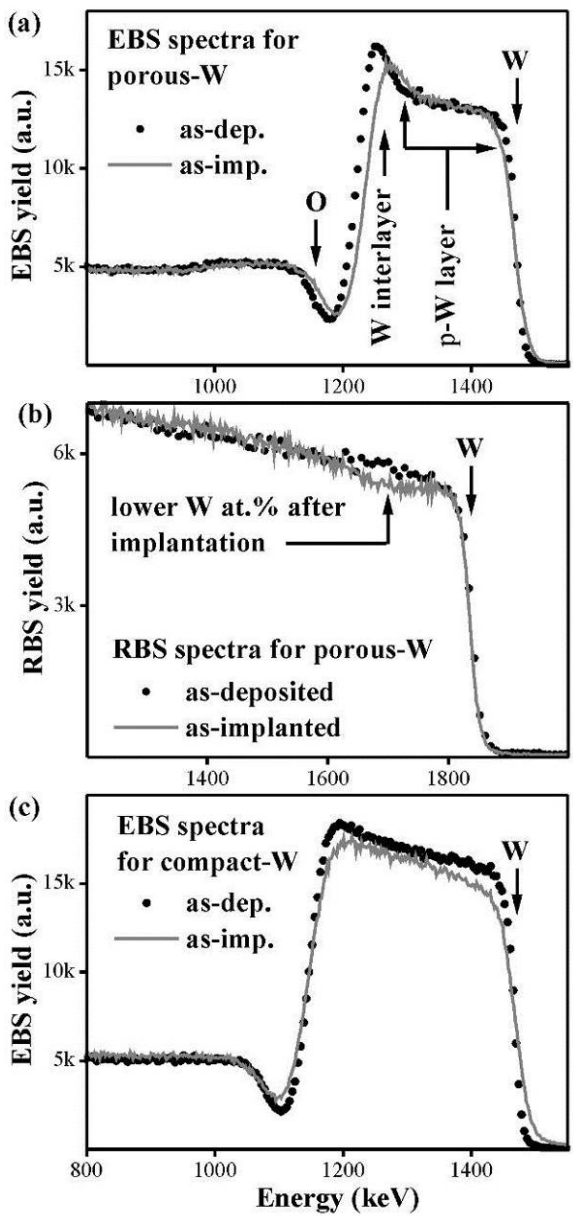

**Fig. 4**

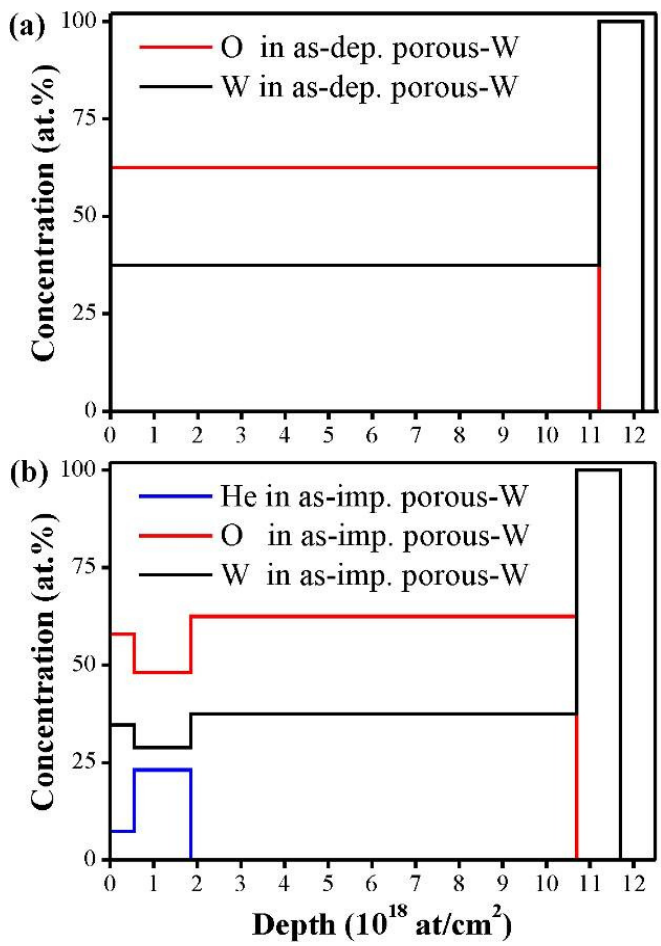

Fig. 5

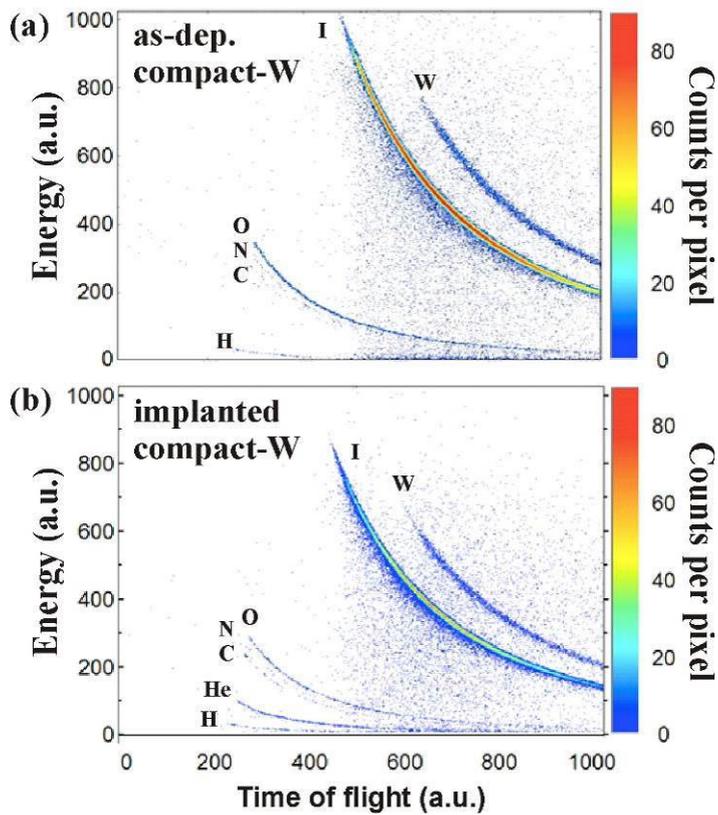

Fig. 6

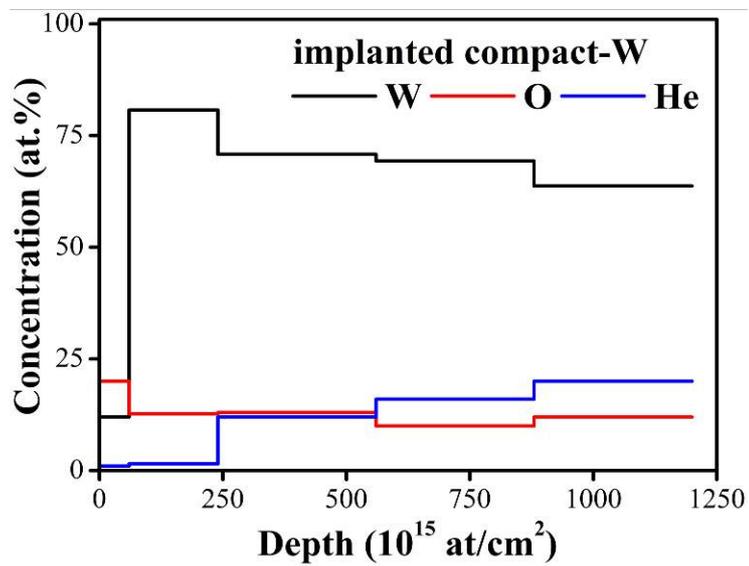

Fig. 7